\documentclass[preprint, prb , showpacs]{revtex4}

\usepackage{graphicx,epsfig} 

\usepackage{dcolumn}

\usepackage{amsmath}

\newcommand{\vect}[1]{\ensuremath{\boldsymbol{#1}}}

\begin{document}

\preprint{}

\title{Non-monotonic angular magnetoresistance in asymmetric spin valves}

\author{Jan \surname Manschot$^{1,2}$}

\author{Arne \surname Brataas$^{2}$}

\author{Gerrit E. W. \surname Bauer$^{1}$}

\affiliation{1: Department of Nanoscience, Delft University of
  Technology, 2628 CJ Delft, The Netherlands\\
2: Department of Physics, Norwegian University of Science and Technology, N-7491 Trondheim, Norway}

\begin{abstract}
The electric resistance of ferromagnet/normal-metal/ferromagnet perpendicular spin valves depends on the relative angle between the magnetization directions. In contrast to common wisdom, this angular magnetoresistance is found to be not necessarily a monotone function of the angle. The parameter dependence of the global resistance minimum at finite angles is studied and the conditions for experimental observation are specified.   
\end{abstract}


\maketitle

The discovery of the giant magnetoresistance (GMR) \cite{baibich} has driven
much of the current research to enrich the functionalities of
electronic circuits and devices employing ferromagnetic elements. The
current perpendicular to plane (CPP) transport
technique\cite{bass,sun1,yang} turned out to be especially suited to
study the physics of spin dependent transport. Nanostructured and
perpendicular spin valves are ideal devices to investigate
the current-induced magnetization reversal,\cite{slonczewski} which
has potential applications for magnetic random access memories. These
structures allow the measurement of the angular magnetoresistance
(aMR)\cite{dauguet,giacomo} introducing an analogue degree of freedom
between the conventional parallel vs. antiparallel digital
configurations. A semiclassical theoretical treatment of the aMR leads
to the concept of a spin-mixing conductance\cite{brataas2} that turned out
useful for phenomena like the spin torque\cite{xia,bauer} and
interface-enhanced Gilbert damping. \cite{tserkovnyak}

This Rapid Communication addresses the aMR of asymmetric perpendicular
spin valves. We show that the parallel magnetization configuration of
ferromagnet(F)/normal metal(N)/ferromagnet heterostructures does not
necessarily correspond to the minimal resistance. This non-monotonic
behavior requires a redefinition of the GMR ratio in terms of the
global maximum and minimum resistances instead of those for parallel
and antiparallel configurations. We discuss how to optimize the
conditions for an experimental observation and demonstrate that the
spin torque is strongly affected by the asymmetry as well. 

First, we summarize necessary concepts from Ref.~\onlinecite{brataas}
for resistive elements such as an interface between a monodomain
ferromagnet with magnetization parallel to the unit vector
$\vect{m}$. The charge and spin current, $I_\mathrm{c}$ and
$\vect{I}_\mathrm{s}$, driven by a potential and spin accumulation
bias, $\Delta\mu_\mathrm{c}$ and $\Delta\vect{\mu}_\mathrm{s}$, read 

\begin{equation}
\label{eq:interfacecharge}
I_\mathrm{c}=\frac{e}{h}\left[(g^{\uparrow\uparrow}+g^{\downarrow\downarrow})\Delta \mu_\mathrm{c}+(g^{\uparrow\uparrow}-g^{\downarrow\downarrow}) \vect{m}\cdot\Delta \vect{\mu}_\mathrm{s}\right],
\end{equation}

\begin{equation}
\label{eq:interfacespin}
\begin{array}{l}
\vect{I}_\mathrm{s}=\frac{1}{4\pi}\vect{m}\left[(g^{\uparrow\uparrow}-g^{\downarrow\downarrow})\Delta \mu_\mathrm{c}+(g^{\uparrow\uparrow}+g^{\downarrow\downarrow})\vect{m}\cdot\Delta 
\vect{\mu}_\mathrm{s}\right] \\
\quad + \frac{1}{4\pi}2\mathrm{Re}(g^{\uparrow\downarrow})\vect{m}\times(\Delta \vect{\mu}_\mathrm{s}\times\vect{m}).
\end{array}
\end{equation}

\noindent where $g^{\uparrow\uparrow}$ and $g^{\downarrow\downarrow}$ are the conductances for electrons
with majority and minority spin, respectively, and $g^{\uparrow\downarrow}$ is the 
mixing conductance for a spin current polarized transverse to the
magnetization. We disregarded $\mathrm{Im}(g^{\uparrow\downarrow})$, which for metallic
interfaces is usually smaller than 10\% of $\mathrm{Re}(g^{\uparrow\downarrow})$.\cite{xia,stiles} It
is convenient to introduce $g=g^{\uparrow\uparrow}+g^{\downarrow\downarrow}$, $p=(g^{\uparrow\uparrow}-g^{\downarrow\downarrow})/g$ and
$\eta=2g^{\uparrow\downarrow}/g$, where $g$ is the total conductance, $p$ the
polarization and $\eta$ the relative mixing conductance.  

\begin{figure}[ht]
\begin{center}
\includegraphics[width=8cm]{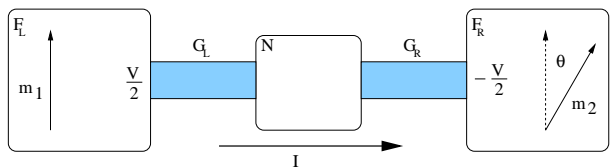}
\caption{Schematic picture of a perpendicular spin valve biased by a
  voltage difference $V$. $\theta$ is the angle between the
  magnetization directions of both reservoirs. The reservoirs and
  contacts need not be identical; the conduction parameters are
  summarized as $G_\mathrm{L}$ and $G_\mathrm{R}$.} 
\label{fig:asmagres}
\end{center}
\end{figure}

Let us examine a two terminal system (F-N-F) as shown in Figure
\ref{fig:asmagres}. The contacts need not be
identical; the conduction parameters are summarized as $G_\mathrm{L}$
and $G_\mathrm{R}$. The electric resistance as function of the angle   
between the magnetization directions of the reservoirs, can simply be
calculated using Eqs. (\ref{eq:interfacecharge}) and
(\ref{eq:interfacespin}), assuming charge and spin conservation on the
normal metal node. For a {\em symmetric} structure
($G_\mathrm{L}=G_\mathrm{R}=G$) the resistance $R(\theta)$
reads:\cite{brataas} 

\begin{equation}
\frac{e^2}{h}R(\theta)=\frac{2}{g}\frac{\tan^2\theta/2+\eta}{(1-p^2)\,\tan^2\theta/2+\eta}.
\end{equation}

\noindent If necessary, spin flip processes in the normal metal can be included.\cite{brataas} A finite angle between the magnetizations causes a spin accumulation on the normal metal node. Since we disregard the imaginary part of the mixing conductance, it lies in the plane of the magnetization vectors. The resistance increases with 
increasing spin accumulation, whose creation costs energy, and thus
with $\theta$. Therefore the resistance is minimal when the
magnetizations are parallel and maximal for $\theta=\pi$. The mixing
conductance can be interpreted as an additional channel for
dissipating the spin accumulation on the normal metal node for
$0<\theta<\pi$; an increasing mixing conductance will therefore
reduce the total resistance. This is the
mechanism behind deviations of the aMR from a simple $\cos^2 \theta/2$
behavior, which can be used to determine the mixing conductance from
experimental curves.\cite{bauer, giacomo} 

In the following, we focus on an asymmetric configuration with $G_\mathrm{L}\neq G_\mathrm{R}$. The asymmetry in conductance
($g_\mathrm{L}\neq g_\mathrm{R}$) causes a charge accumulation on the
normal metal node. Similarly, when $p_\mathrm{L}\neq p_\mathrm{R}$, a 
spin accumulation is excited on the normal metal node even for
$\theta=0$. We find here that configurations with $\theta\neq
0$ may correspond to a spin accumulation that is smaller than that of
the parallel one, and therefore a global resistance minimum at finite
angles. The recipe for a significant effect is a large polarization of the current by the source contact (e.g. $p_\mathrm{L}\approx 1$)  and efficient
dissipation of the spin accumulation for a finite angle $\theta$ by a
large mixing conductance ($\eta_\mathrm{R}>1$) of the drain. The
polarization direction of the spin current differs from the
magnetization directions for finite angles $\theta$ as in the
symmetric case, but the noted asymmetry forces it to be close to the
magnetization direction of the source contact. A large mixing
conductance $g^{\uparrow\downarrow}$ favors the transverse 
over the longitudinal spin current. Spins on the
normal metal node therefore escape easily and the reduced spin accumulation 
is equivalent to a decrease of the total resistance. This interplay between spin accumulation and magnetization angles strongly modifies the total aMR profile.

Unfortunately, the exact equations for $R(\theta)$ are not very transparent. 
A perturbation approach to these equations is not helpful because no small parameters can be identified for the experimentally relevant metallic structures. However, we did find relatively simple analytical expressions for the angle $\theta_\mathrm{m}$ of the global resistance minimum as well as a simple expression for the maximal aMR in the limit $\eta_\mathrm{R}\gg1$. 

We derive that next to $\theta=0,\,\pi$, the resistance may have
extrema at two additional angles:

\begin{eqnarray}
\label{eq:solution1}
\cos \theta_\mathrm{m1}=\left(\frac{p_\mathrm{R}}{p_\mathrm{L}}\right)\left(\frac{1+\left(\frac{g_\mathrm{L}}{g_\mathrm{R}}\right)
\frac{1-p_\mathrm{L}^2}{\eta_R}} {1-\frac{1-p_\mathrm{R}^2}{\eta_R}}\right), \\
\label{eq:solution2}
\cos \theta_\mathrm{m2}=\left(\frac{p_\mathrm{L}}{p_\mathrm{R}}\right)\left(\frac{1+\left(\frac{g_\mathrm{R}}{g_\mathrm{L}}\right)\frac{1-p_\mathrm{R}^2}{\eta_L}}
{1-\frac{1-p_\mathrm{L}^2}{\eta_L}}\right).
\end{eqnarray}

\noindent where the absolute value of $\cos\theta_\mathrm{m}$ must be
smaller than unity, which is clearly not the case for a symmetrical
spin valve. The condition for one extra extremum is easily
fulfilled. {\em Two } additional extrema are not consistent with the
condition $\eta_\mathrm{L},\eta_\mathrm{R}>1$, which rigorously holds
for high contact resistances,\cite{brataas} but not necessarily for
highly transparent interfaces.\cite{bauer} It can be proven that when one
extremum exists and $\eta_\mathrm{L},\eta_\mathrm{R}>1$, the extremum
is the global minimum and located in the interval $0<\theta<\pi/2$. When
$\eta_\mathrm{R}<1-p_\mathrm{R}^2$ (that does not seem very likely
for metals), an additional extremum may exist in the interval $\pi/2<\theta<\pi$.
It turns out to be a maximum that can be understood in the same way as the minimum.
 Additional minima and maxima may even coexist for specific parameter combinations, which do not appear relevant for metallic spin valves, however. 

The position of the global minimum does not depend on $\eta$ of the
source contact. The source polarizes the current through
the total structure parallel to its magnetization, therefore the
source $\eta$ does not play a role at all. The component of
the spin current orthogonal to the magnetization is called the spin
torque\cite{xia} acting on this magnetization, since it is absorbed
by the magnetic order parameter and may excite the
magnetization when exceeding a threshold value.\cite{slonczewski,kent,kiselev} In the global minimum
the spin torque on the source magnetization vanishes with the transverse
component of the spin current. The spin torque on the drain
is large, but not at the maximum as a function of $\theta $. 

Let us choose the left lead to be the polarizing source $(p_\mathrm{L}>p_\mathrm{R})$ and the right lead to be the dissipating drain $(\eta_\mathrm{L}>\eta_\mathrm{R})$. The condition for a non-collinear resistance minimum is now: 

\begin{equation}
\label{eq:condition}
\left| \left(\frac{p_\mathrm{R}}{p_\mathrm{L}}\right)\left(\frac{1+\left(\frac{g_\mathrm{L}}{g_\mathrm{R}}\right)
\frac{1-p_\mathrm{L}^2}{\eta_R}} {1-\frac{1-p_\mathrm{R}^2}{\eta_R}}\right) \right|<1.
\end{equation}

\noindent When the second factor is larger than one (true for
$\eta_\mathrm{R}>1$), only a polarization ratio
$p_\mathrm{L}/p_\mathrm{R} > 1$ can save this inequality. The condition
is never fulfilled when the left hand side of Equation
(\ref{eq:condition}) diverges: 

\begin{equation}
\frac{1-p_\mathrm{R}^2}{\eta_\mathrm{R}}=\frac{2\left(\frac{1}{g^{\uparrow\uparrow}_\mathrm{R}}
+\frac{1}{g^{\downarrow\downarrow}_\mathrm{R}}\right)^{-1}}{g^{\uparrow\downarrow}_\mathrm{R}}\approx 1.
\end{equation}

\noindent Therefore $g^{\uparrow\downarrow}_\mathrm{R}$ should be considerably larger than  $g^{\uparrow\uparrow}_\mathrm{R}$ and $g^{\downarrow\downarrow}_\mathrm{R}$. When the average conductance of 
the source is smaller than the mixing conductance of the
drain, the numerator

\begin{equation}
\left(\frac{g_\mathrm{L}}{g_\mathrm{R}}\right)\frac{1-p_\mathrm{L}^2}{\eta_R}=
\frac{2\left(\frac{1}{g^{\uparrow\uparrow}_\mathrm{L}}+\frac{1}{g^{\downarrow\downarrow}_\mathrm{L}}\right)^{-1}}
{g^{\uparrow\downarrow}_\mathrm{R}},
\end{equation}

\noindent reduces $\cos \theta_\mathrm{m}$, and hence increases $\theta_\mathrm{m}$.

The GMR ratio is usually defined in terms of the resistance in
parallel or antiparallel configurations, in terms of the previously
introduced parameters: 

\begin{eqnarray}
\label{eq:mr*}
\lefteqn{G\!M\!R^*=\frac{R^\mathrm{ap}-R^\mathrm{p}}{R^\mathrm{ap}} =} \nonumber \\
& & \frac{g_\mathrm{L}g_\mathrm{R}\left[(p_\mathrm{L}+p_\mathrm{R})^2-
(p_\mathrm{L}-p_\mathrm{R})^2\right]}{(g_\mathrm{L}+g_\mathrm{R})^2-(g_\mathrm{L}p_\mathrm{L}-g_\mathrm{R}p_\mathrm{R})^2},
\end{eqnarray}

\noindent where $R^\mathrm{ap}$ and $R^\mathrm{p}$ are the resistances
for antiparallel and parallel configuration, respectively. Hence, the
GMR ratio increases when the total polarization increases, as
expected. However, when the difference between the polarizations of
both sides is large, $G\!M\!R^*$ decreases, because of a larger spin
accumulation on the normal metal node for $\theta=0$, as noted
above. Since now $R^\mathrm{ap}-R^\mathrm{p}$ is no longer the maximal
resistance difference, a new definition for the magnetoresistance is
appropriate in terms of the global maximum (for which we still take
the antiparallel configuration) and the newly found global minimum. In
the limit of large $\eta_\mathrm{R}\gg1$ we arrive at the simple
result

\begin{equation}
\label{eq:mr}
G\!M\!R=\frac{R^\mathrm{ap}-R^\mathrm{m}}{R^\mathrm{ap}}=G\!M\!R^* \frac{(p_\mathrm{L}+p_\mathrm{R})^2}{4p_\mathrm{L}p_\mathrm{R}},
\end{equation}

\noindent where $R^\mathrm{m}$ is the global minimum of the
resistance. It can easily be verified that $G\!M\!R$ is indeed larger
than $G\!M\!R^*$. 

Next we investigate the conditions under which this enhanced
magnetoresistance can be measured in magnetic spin valves with a
current perpendicular to plane geometry.\cite{bass} Even for identical
magnetic layers an asymmetry can be realized by a spin independent
resistance or tunnel barrier at the outside of one of the magnetic
films, as long as the spin diffusion length is larger than the total
bilayer. Such an additional series resistor then effectively decreases
$p$, $g^{\uparrow\uparrow}$ and $g^{\downarrow\downarrow}$ of this magnet. Because the spin current
normal to the magnetization is absorbed by the
magnet\cite{slonczewski} over a couple of monolayers, $g^{\uparrow\downarrow}$ is not
modified by the extra resistance.\cite{xia} $g^{\uparrow\downarrow}$ thus can indeed be
engineered to be larger than $g^{\uparrow\uparrow}$ for a given contact such that a
non-monotonic aMR can be expected.  

Spin dependent bulk resistances contribute to the aMR over
thicknesses smaller than the spin diffusion length. Copper and cobalt have
relatively large spin flip lengths, respectively $250\,\mathrm{nm}$
and $50\,\mathrm{nm}$, which makes them useful materials to
explore this effect. $\mathrm{Al}_2\mathrm{O}_3$ tunnel barriers are
routinely used for tunnel MR studies and suitable materials for the
present purposes.

\begin{figure}[ht]
\begin{center}
\includegraphics[width=8cm]{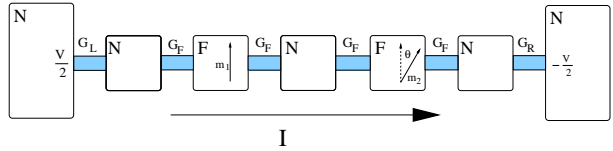}
\caption{Schematical picture of a thin film pillar with two ferromagnetic (F), three normal metal (N) layers and two normal metal reservoirs.}
\label{fig:pillar}
\end{center}
\end{figure}

The full aMR profile can best be calculated numerically. We consider a structure consisting of two identical cobalt layers
(thickness is $3\,\mathrm{nm}$) separated by a thin copper layer
($10\,\mathrm{nm}$). Both magnets are sandwiched by tunnel junctions,
another copper layer, and finally normal metal reservoirs as sketched
in Fig. \ref{fig:pillar}. Bulk resistances of copper and cobalt are
disregarded because they are relatively very small. $G_\mathrm{F}$
symbolizes all 
conductance parameters of a copper-cobalt interface, $G_\mathrm{L}$
and $G_\mathrm{R}$ stand for the outer normal resistances including
the tunnel junctions. For interfaces between a ferromagnet and a
normal metal, $g^{\uparrow\downarrow}$ ( $\sim$ number of modes in the normal metal)
usually lies between $g^{\uparrow\uparrow}$ and 
$g^{\downarrow\downarrow}$. For a Co/Cu interface $g$ is typically $1413\times10^{3}$
(for an interface cross section of $140\times90\,\mathrm{nm}^2$),
$p=0.75$ (Ref. \onlinecite{yang}) and $\eta=0.38$ (Ref. \onlinecite{xia}); these values include
the Boltzmann corrections for transparent interfaces.\cite{bauer} In
order to compare configurations with different values for
$G_\mathrm{L}$ and $G_\mathrm{R}$, its series resistance is assumed
constant at $1/G_\mathrm{L}+1/G_\mathrm{R}=0.37\,\Omega$.  

\begin{figure}[ht] 
\begin{center}
\includegraphics[width=8cm]{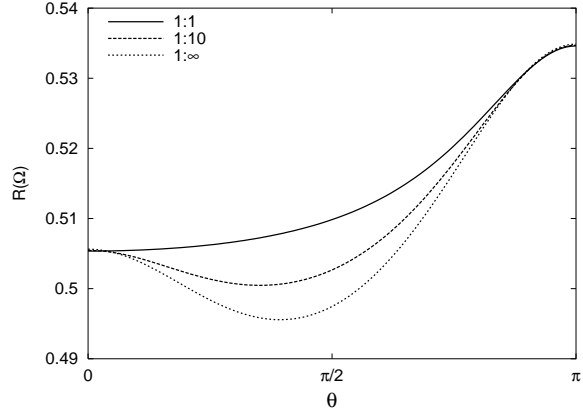}
\caption{Angular dependence of the thin film pillar resistance. $\theta$ is the angle between the magnetization directions of both layers.}
\label{fig:agmr}
\end{center}
\end{figure}

The computed aMR is presented in Figure \ref{fig:agmr} for three
different ratios $G_\mathrm{L}:G_\mathrm{R}$. When $G_\mathrm{L}$ differs sufficiently from $G_\mathrm{R}$, the
global minimum shifts away from the parallel configuration, as
predicted. The position of the global minimum, $\theta_\mathrm{m}$
increases with increasing polarization 
contrast. The GMR ratio increases as well, which is in qualitative
agreement with Eq. (\ref{eq:mr}).  

Finally we compute the spin torque, $i.e.$ the transverse component of
the spin current. The spin torque in spin
valves is governed by similar expression as the charge
current\cite{bauer} and is strongly affected by the asymmetry as
well. It is convenient to normalize the spin torque by the charge
current: 

\begin{equation}
i_\mathrm{s}=\frac{|\vect{m}\times(\vect{I}_\mathrm{s}\times\vect{m})|}{|I_\mathrm{c}|}.
\end{equation}

\begin{figure}[ht]
\begin{center} 
\includegraphics[width=8cm]{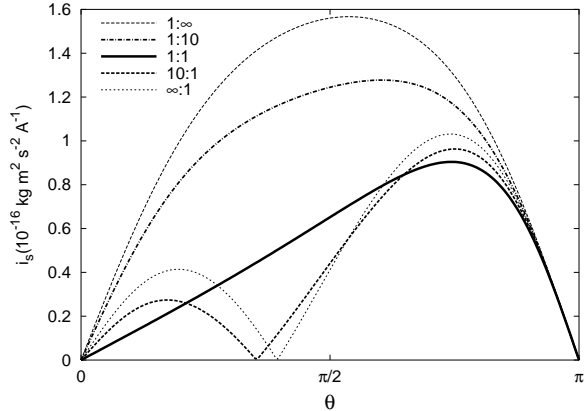}
\caption{Normalized spin torque on the left magnetization as function
 of $\theta$.}
\label{fig:spintorque}
\end{center}
\end{figure}

\noindent In Figure \ref{fig:spintorque}, $i_\mathrm{s}$ of the left
magnetization is plotted as a function of $\theta$ and different $G_\mathrm{L}/G_\mathrm{R}$ ratios. The zeroes in the intervals $0<\theta<\pi/2$
illustrate that when the left side is the polarizing source, the spin
torque at the global minimum vanishes, which agrees with the finding
that the resistance minimum is not a function of $\eta_\mathrm{L}$. We
observe that the spin torque is strongly enhanced when
$G_\mathrm{L}/G_\mathrm{R}\rightarrow 0$ because the relative mixing conductance of
the left hand side is then highly increased, which physically means
that the spin accumulation on the normal metal node easily can be dissipated.  

To summarize, we have shown that the angular magnetoresistance (aMR)
of perpendicular spin valves can be a non-monotonic function when the
contacts between the central normal metal node and the outer
ferromagnets differ. An analytical expression is derived for the angle
$\theta_\mathrm{m}$ at which the magnetoresistance has its global
minimum. A new definition for the GMR ratio is proposed to take this
effect into account. This GMR ratio is now larger than the
conventional definition in terms of the resistance of parallel and
antiparallel configurations. The spin torque in asymmetric structures
is also importantly modified.

J.M. thanks Martin Gr{\o}nsleth and Jan Petter Morten for fruitful
discussions. This work has been supported by ERASMUS, the FOM and the
NEDO joint research program ``Nano-Scale Magnetoelectronics".

\bibliographystyle{apsrev}

\end{document}